\documentclass[fleqn,twoside]{article}
\usepackage{espcrc2}
\usepackage{graphicx}
\usepackage[figuresright]{rotating}

\newcommand{\sss}{\scriptscriptstyle}
\def\lg{^{\rm long}}
\def\Amix{\mathcal{A}_{\rm mix}}
\def\Adir{\mathcal{A}_{\rm dir}}

\def \apl{A_{\| L}}
\def \apr{A_{\| R}}
\def \appl{A_{\bot L}}
\def \appr{A_{\bot R}}
\def \al{A_0}
\def \ap{A_{\|}}
\def \app{{A}_{\bot}}

\newcommand{\AmS}{{\protect\the\textfont2
  A\kern-.1667em\lower.5ex\hbox{M}\kern-.125emS}}

\title{Extracting the $B_s-{\bar B_s}$ mixing angle from $B \to VV$ decays and comments on the puzzling $B\to K^* l^+l^- $ decay}

\author{Joaquim Matias\address[MCSD]{IFAE \& Universitat Aut\`onoma de Barcelona,
        E-08193 Bellaterra, Barcelona, Spain}}

\begin{document}

\begin{abstract}
Three different strategies  to extract the weak mixing phase $\phi_s$ of the $B_s$ system using $B \to VV$ decays ($B_{s} \to K^{*0} {\bar K^{*0}}$, $B_s \to \phi {\bar K^{*0}}$ and $B_s \to \phi\phi $) are discussed. Those penguin-mediated decays are computed in the framework of a new combined QCD-Factorisation/Flavour Symmetry Method. Also some comments on the recent interesting results found by Babar concerning the decay $B \to K^* l^+l^- $ are included.
\vspace{1pc}
\end{abstract}

\maketitle

\section{Introduction}
The present era of Precision Flavour Physics may turn soon due to the excellent performance of the $B$-factories and Tevatron (and hopefully  LHC)  into an era of  Precision Flavour {\bf New} Physics. Rare $b \to s$ transitions in several inclusive and exclusive modes start exhibiting some tension with the SM predictions\cite{silver}. Indeed, there is certain theoretical prejudist\cite{prejud} to expect deviations in $b \to s$ transitions and SM-like results in the corresponding $b \to d$ ones.

In order to reach the accuracy needed in some $B$ decay processes to get to the  discovery level it is of an utmost importance to refine the me-thods used to predict those decays. There are two main approaches in the literature: $1/m_b$-expansion based methods, namely, QCD Factorization (QCDF)\cite{BBNS,BN}, Soft collinear effective theories \cite{scet} or PQCD\cite{pqcd} and Flavour Symmetry methods like U-spin \cite{uspin}. Here I will discuss a new method \cite{dmv} that combines the predictive power of QCDF techniques with the model-independence of Flavour symmetries. Moreover, a main advantage of this method is that it reduces substantially the sensitivity to the dangerous chirally enhanced IR divergences in QCDF and the arbitrariness in the choice of the size of SU(3) breaking in Flavour Symmetries.

Before briefly discussing the method and its application to decays of $B$ mesons into vectors \cite{dmv2} to extract the weak mixing angle $\phi_s$, I will open a parenthesis to comment on the results of the measurement of the $B \to K^* l^+l^-$ decay. Here I will focus on certain observables that were proposed in \cite{fm,km,lm}.

\section{Comments on $B \to K^* l^+ l^-$}

Recently, Babar has
found a set of very interesting results \cite{prlexp,kevin} while measuring observables based on
the $B \to K^* l^+ l^-$ channel, namely: i) longitudinal polarization fraction of the $K^*$: $F_L$, ii) Forward-Backward asymmetry: $A_{FB}$, iii) isospin asymmetry $A_I$ of $B^0\to K^{*0}l^+l^-$ and $B^\pm \to K^{*\pm} l^+l^-$ channels and soon iv) the transverse asymmetry: $A_T^{(2)}$.
I will discuss them in turn and comment on the impact of  two scenarios (flipped sign solution for $C_7^{eff}$) and Right Handed currents (RH) scenario ($C_7^{eff\prime}\neq 0$). Even if, at a first sight, the flipped sign scenario  (not my favorite one) may appear as a possible solution for all these deviations, indeed it is not a completely satisfactory solution. Moreover, this scenario is somehow disfavoured (under certain assumptions) by the inclusive mode, but not ruled out. Of course, this should be taken only as an exercise, since data should still give the final answer and,  other scenarios may led to a better explanation if data changes substantially.
\begin{itemize}
\item $F_L(s)=\frac{|{A}_0|^2}{|{ A}_0|^2 + |{A}_{\|}|^2
+ |A_\perp|^2}$ \cite{km}, where $A_{0,\perp,\|}$ are, respectively, the longitudinal, perpendicular and parallel spin amplitude of the $K^*$. The theoretical prediction for this observable was computed in QCDF at NLO  in \cite{km} and including possible $\Lambda/m_b$ corrections of order 10\% in each amplitude in \cite{lm}. From \cite{lm} (see Fig.1) one gets an average value (weighted by the distribution $d\Gamma/dq^2$) in the region $1\le q^2 \le 6.25 {\rm GeV}^2$ of $0.83 \pm 0.08$ ($\xi_{\perp}(0)$ was taken 0.35). An update of this calculation has been done in \cite{newpaper} with an average value $0.86 \pm 0.05$ slightly higher due to the different choice in \cite{newpaper} for $\xi_{\perp}(0)=0.26$. Recently an experimental averaged value of $0.35 \pm 0.16$ \cite{prlexp} was measured on a larger region $4 m_\mu^2 \le q^2 \le 6.25 {\rm GeV}^2$.
\begin{figure*}[ht]
\begin{center}
\includegraphics[width=14pc]{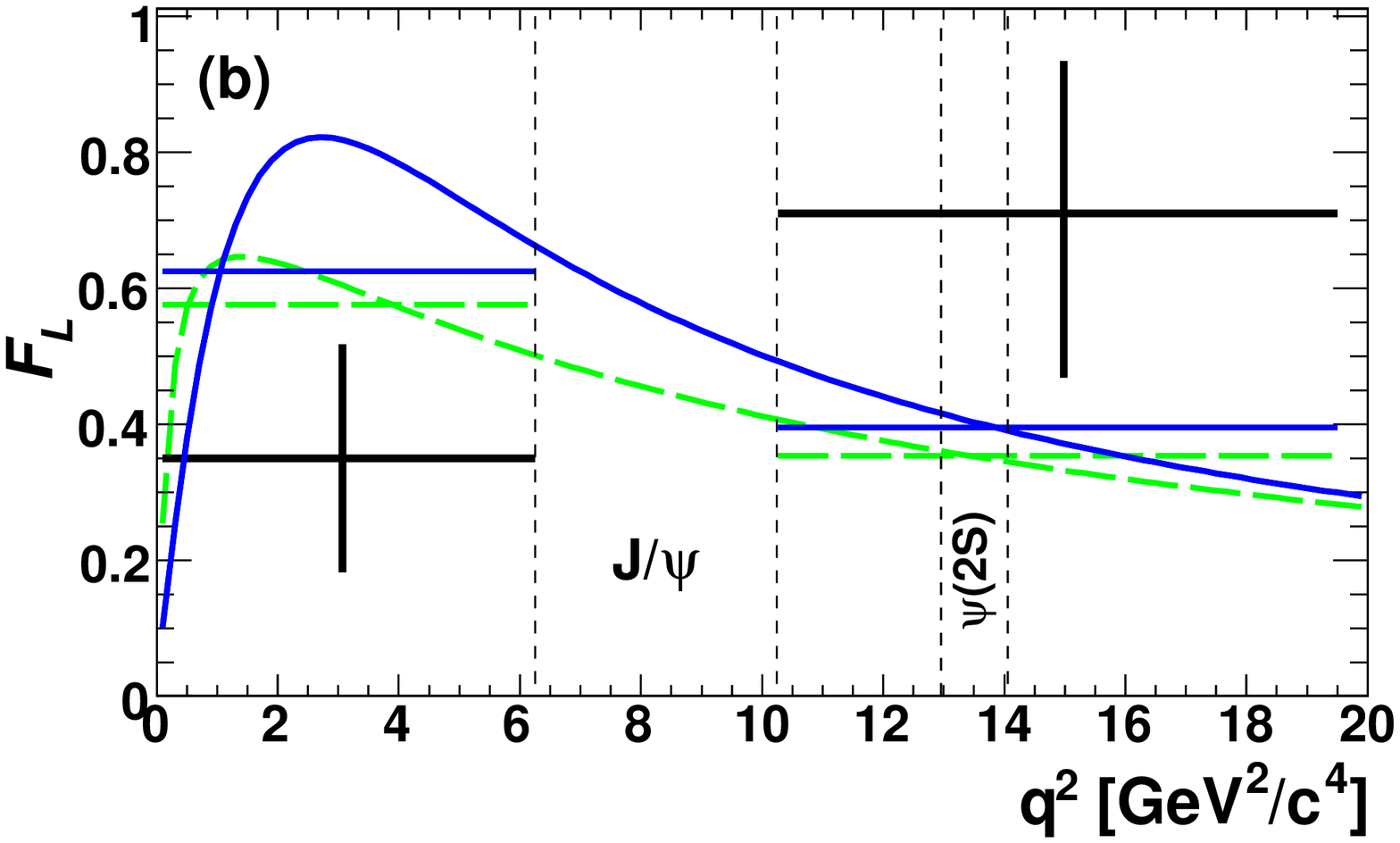}
\includegraphics[height=8pc,width=13.8pc]{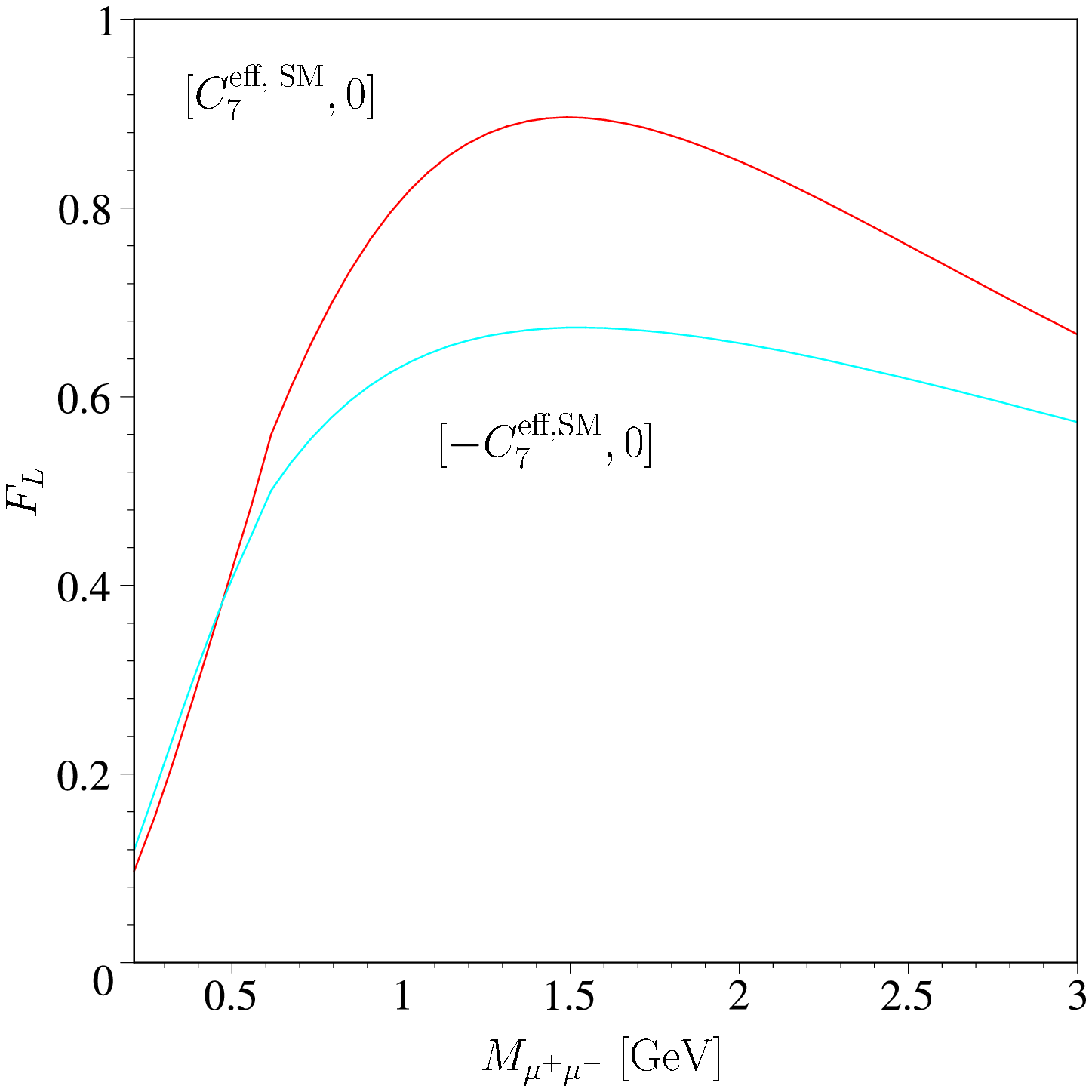} \end{center}
\begin{center}
\hspace*{-0.4cm}\includegraphics[height=8pc,width=16pc]{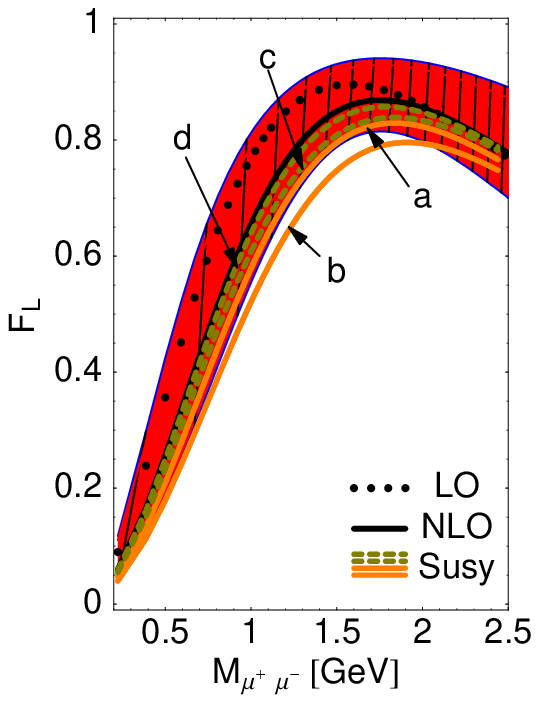}
\vspace*{1cm}{\includegraphics[height=8.8pc,width=13.8pc,bb=10 20 300 220,clip]{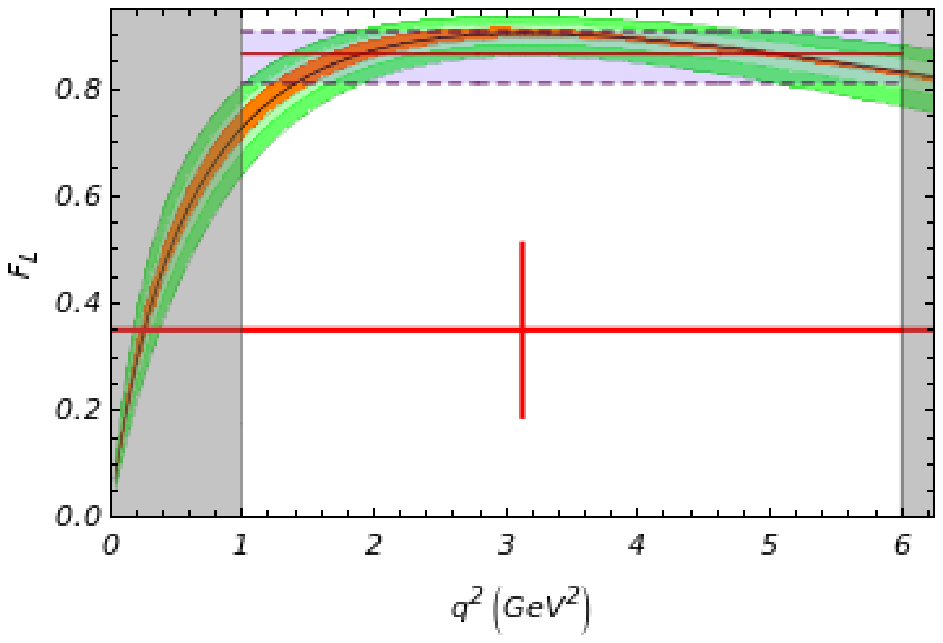}} \end{center}
\vspace*{-1.5cm}
\caption{Top left: Experimental result on $F_L$ \cite{prlexp}. Top right: Flipped sign solution $C_7 ^{{eff}}$ \cite{km}.
Bottom left:
Theoretical calculation of $F_L$ including $\Lambda/m_b$ with $\xi_{\perp}(0)=0.35$ \cite{lm}. Bottom right: Updated theory average\cite{newpaper} (using $\xi_{\perp}(0)=0.26$) in $1\le q^2 \le 6.25 {\rm GeV}^2$ region (horizontal band) and experimental cross average in the larger region $4 m_\mu^2 \le q^2 \le 6.25 {\rm GeV}^2$.}
\label{fig:largenenough}
\end{figure*}
Even if these two numbers
cannot be yet compared, it is presumable that the averaged experimental value once taken in the $1\le q^2 \le 6.25 {\rm GeV}^2$ region will certainly increase but still probably will remain a low value.
(One can do the opposite exercise and compute the theory average from $4 m_\mu^2$ to $6.25 {\rm GeV}^2$, not taken into account possible low resonances effects obtaining $0.67\pm 0.08$. Of course, this number should be taken {\it only} as an indication). Let me now discuss the impact of the two scenarios: first, it was found in \cite{lm} that supersymmetry with  non-minimal
flavour violation in the down sector, as an example of RH current contribution ($C_7^{eff\prime}\neq0$) do not deviate much from the SM
region (Fig.1). Indeed, beyond the supersymmetric case, picking up some extreme but allowed values of $C_7^{eff\prime}$ one finds that it is hard to go down more than a 10\% with respect to the SM band. The other scenario,  the flipped sign solution for $C_7^{eff}$ deviate more substantially\cite{km} (see Fig.1) from the SM region and in the direction of the measured valued.

\item
$
A_{\rm FB}(s)=\frac{3}{2}\frac{\Re(\apl\appl^*) - \Re(\apr\appr^*)}
    {|\al|^2 + |\ap|^2 + |\app|^2}
 $
 \cite{fm,afbsm}. The forward-backward asymmetry, is particularly interesting on  its zero,
where form factors drop  at leading order  giving a precise relation between $C_7^{eff}$ and $C_9$, but also due to its sensitivity to the sign of $C_7^{eff}$. Experimentally, it was found (Fig.2) again a deviation that tend to prefer the reversed sign of $C_7^{eff}$. An scenario compatible with this situation was discussed in \cite{fm} in the case of MSSM with large $\rm tan \beta$ (see Fig.2). Again for this observable, like in the case of $F_L$, RH currents originating from a supersymmetric model do not seem to deviate substantially from the SM prediction.

\item $A_I=\frac{{d\Gamma[B^0\to K^{\ast0}\ell^+ \ell^-]}/{ds} -
{d\Gamma[B^\pm\to  K^{\ast\pm}\ell^+ \ell^-]}/{ds}}
{{d\Gamma[ B^0\to K^{\ast0}\ell^+ \ell^-]}/{ds} +
{d\Gamma[B^\pm \to  K^{\ast\pm}\ell^+ \ell^-]}/{ds}}
\label{dAI}
$ \cite{fm}. This asymmetry
in the SM arises from graphs where a photon is radiated from the spectator
 quark in annihilation or spectator-scattering diagrams. The sensitivity to
the different charge of the spectator quark for a $B^0$ or a $B^+$ induces
a non-zero value. The computation of this isospin
 asymmetry for $B \to K^* l^+ l^-$ in the framework of QCDF
was done in \cite{fm}.
It was found there that for values of $q^2>0$ no sizeable isospin asymmetry is expected
in the SM (see Fig.3).   The case of $q^2=0$ was computed in \cite{nk}. In the limit $q^2 \to 0$, where the photon pole dominates, our isospin asymmetry
reduces to $A_I(B \to K^* \gamma)={\rm Re}(
b_d^\perp(0)-b_u^\perp(0))$ (in agreement with \cite{nk}). It is remarkable
the good agreement with the SM prediction and the experiment
at this point and this is puzzling. The $q^2=0$ solution is also sensitive to the sign of $C_7$, although relative to the sign of $C_5-C_6$.
On the contrary, for larger values of $q^2$ the isospin asymmetry is
dominated by the longitudinal polarization amplitude and the dominant operators are $O_3-O_4$. So even if one may devise a solution to accommodate SM at $q^2=0$ and beyond SM at $q^2>0$ any solution looks a bit unnatural.  Remarkably again it is where the longitudinal polarization dominates where
it was found experimentally (Fig.3) a  deviation from the SM prediction. Barring the $q^2=0$ problem, in \cite{fm} it was found that
for the scenario of MSSM with large $\tan\beta$ the flipped sign solution of $C_7^{eff}$ induces negative values for the isospin asymmetry (Fig.3), pointing towards the experimental result, although still far from the measured value.
\item $
A^{(2)}_{\rm T}({s})=\frac{|\app|^2 -
|\ap|^2}{|\app|^2 + |\ap|^2}
$ \cite{km,lm}. This last observable, still not measured could provide an important piece of the puzzle. This is an observable constructed to minimize theoretical uncertainties and show a maximal sensitivity to the presence of RH currents.
If all the impact of New Physics consist on flipping the sign of $C_7^{eff}$ one should not see any deviation from the SM prediction in $A^{(2)}_T$.  
On the contrary, if it deviates, the presence of RH currents would be favored (with or without flipped $C_7^{eff}$).

\end{itemize}

To conclude, we observe that the solution with the flipped sign of $C_7^{eff}$ seems to go in the right direction (but it is still not sufficient) to explain the preliminary observed deviations. However, it has two important caveats: it is disfavoured by the inclusive mode and, moreover, it may require a weird solution to avoid conflicts with $A_I$ at $q^2=0$. RH current solution ($C_7^{eff \prime}\neq0$) seems not to deviate enough from SM prediction for $F_L$ and $A_{FB}$ ($A_T^{(2)}$ may help to favour or rule out this solution). Second, observables containing the longitudinal polarization
($F_L$, $A_{FB}$, $A_I$) seems to systematically exhibit deviations (consistently with this remark one would expect that $A_T^{(2)}$, like $A_I$ at $q^2=0$ will not deviate). In this sense,
the new longitudinal observables proposed in \cite{newpaper} may play an important role.
Finally, it will be very interesting
to see the comparison between theory and experiment strictly inside the theoretically well controlled region $1<q^2<6.25 {\rm GeV}^2$. This closes the parenthesis.

%
\begin{figure}[htb]
\includegraphics[width=15pc]{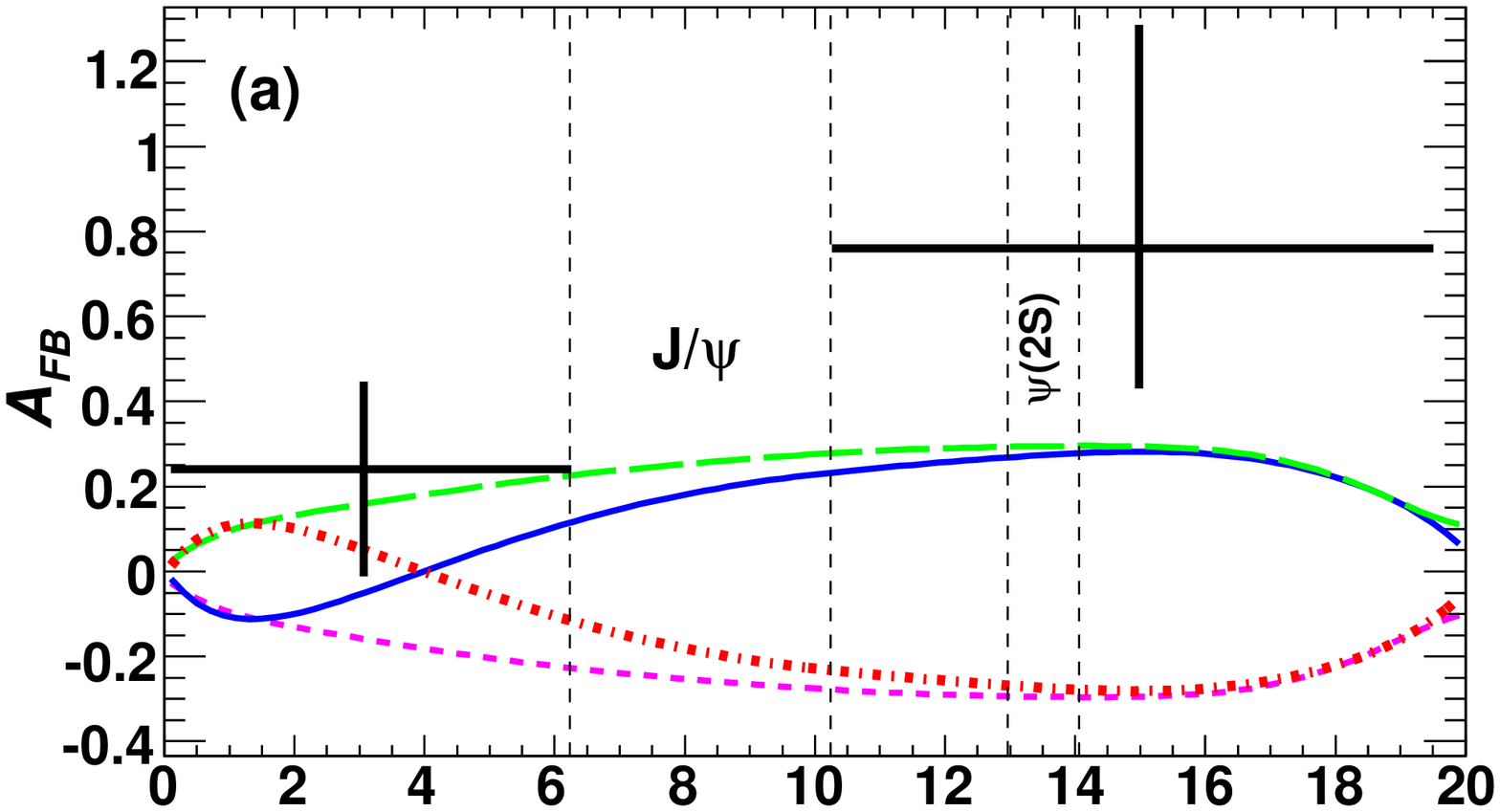}
\vspace*{1cm}
\hspace*{-0.4cm}\includegraphics[width=15.5pc]{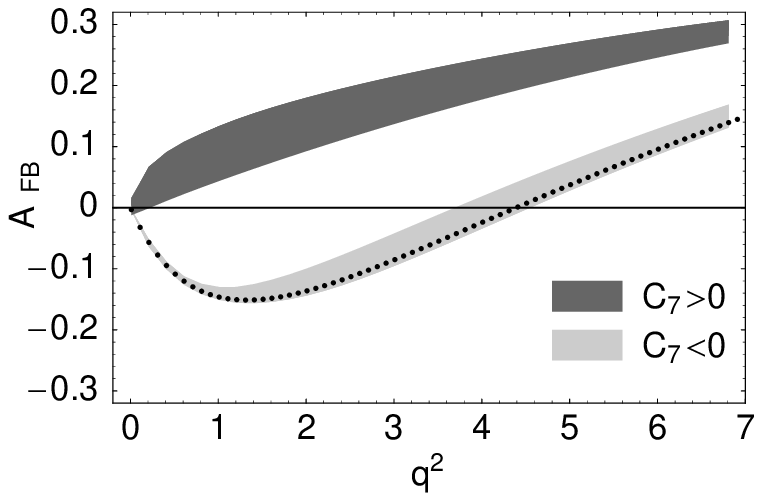}
\vspace*{-1.5cm}
\caption{Top: Experimental forward-backward asymmetry\cite{prlexp}. Bottom: Theoretical prediction for SM and MSSM with large $\tan\beta$\cite{fm} (dark region: flipped $C_7^{eff}$).}
\label{fig:largenenough}
\end{figure}
\begin{figure}[htb]
\vspace*{-0.2cm}
\includegraphics[height=10pc,width=14pc]{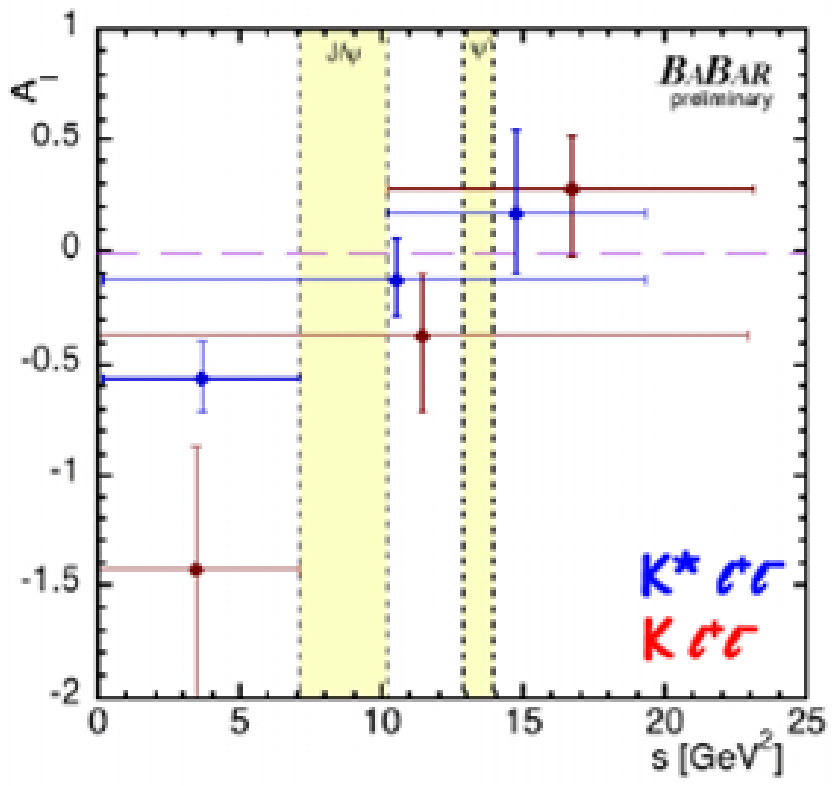}
\hspace*{0.0cm}\includegraphics[width=17pc,bb=114 390 300 
501,clip]{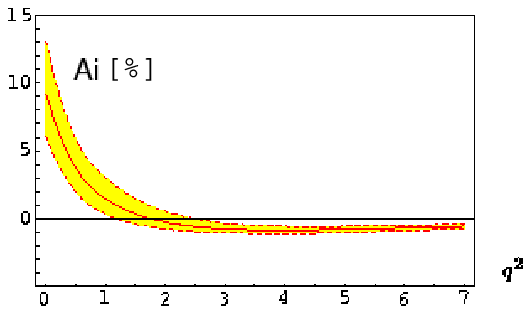}
\hspace*{0.0cm}\includegraphics[width=14pc,bb=34 14 241 151,clip]{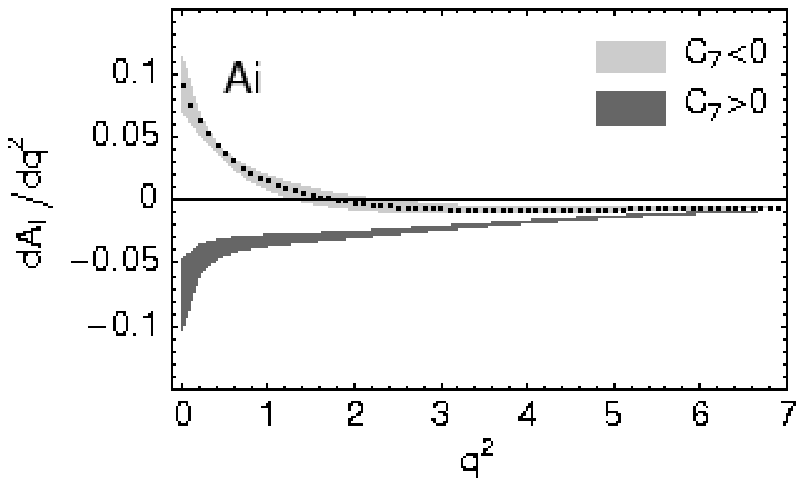}
\vspace*{-1cm}
\caption{From top to bottom: i) Experimental measurement of $A_I$\cite{kevin}, ii) Theoretical prediction for $A_I$ in the SM \cite{fm}, iii) MSSM prediction for large $\tan\beta$ (dark region: flipped $C_7^{eff}$)\cite{fm}.}
\label{fig:largenenough}
\end{figure}
\begin{figure}[htb]
\vspace*{-0.5cm}
\includegraphics[height=13pc,width=15pc]{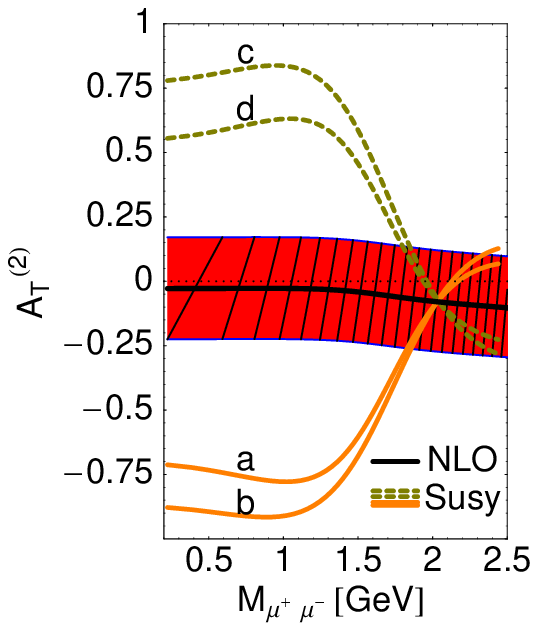}
\vspace*{-1cm}
\caption{Theoretical prediction for $A_T^{(2)}$ in SM and supersymmetry \cite{lm}}
\label{fig:largenenough}
\end{figure}


\section{Description of the Method}

\noindent One of the main source of uncertainties in QCDF comes from IR divergences originating from:

$\bullet$ Hard spectator-scattering: Hard gluons exchange
between spectator quark and the outgoing energetic meson gives
rise to integrals of the following type  (see \cite{BN} for definitions):
\begin{eqnarray}
   H_i(M_1M_2)
   &=& C
   \int_0^1\!dx \int_0^1\!dy \left[
   \frac{\Phi_{M_2}(x)\Phi_{M_1}(y)}{\bar x\bar y}\right.\cr
   &+& \left. r_\chi^{M_1}\,\frac{\Phi_{M_2}(x)\Phi_{m_1}(y)}{x\bar y} \right], \nonumber
\end{eqnarray}
where the second term (formally of order $\Lambda/m_b$) diverges
when $y\to 1$.

$\bullet$ Weak annihilation: These type of diagrams also
exhibit endpoint IR divergences as it is explicit in the corresponding integrals:
\begin{eqnarray} A_1^i =
\pi\alpha_s \!\!\!\!\!\!\!\!\!&&\int_0^1 dx dy\,
    \left\{  \Phi_{M_2}(x)\,\Phi_{M_1}(y)
    \left[ \frac{1}{y(1-x\bar y)} \right. \right.
     \nonumber \cr &+& \left. \left.
    \frac{1}{\bar x^2 y} \right]
    + r_\chi^{M_1} r_\chi^{M_2}\,\Phi_{m_2}(x)\,\Phi_{m_1}(y)\,
     \frac{2}{\bar x y} \right\}. \end{eqnarray}
Both divergences are modeled in the same way in QCDF: %
$
   \int_0^1\frac{d y}{\bar y}\Phi_{m_1}(y)
  \equiv \Phi_{m_1}(1)\,X_{H,A}^{M_1}
    +  {r,} $ with $r$ a finite piece and $X_{H,A}=(1+
\rho_{H,A})\, {\rm ln} (m_b/\Lambda)$. These divergences
are the main source of error in QCDF.
If one splits the SM amplitude for a $B$-decay into two mesons in two pieces:
$
\bar{A}\equiv A(\bar{B}_q\to M \bar{M})
  =\lambda_u^{(q)} T_M^{qC} + \lambda_c^{(q)} P_M^{qC}\,,
$ with $C$ denoting the charge of the decay products, and $\lambda$'s the
products of CKM factors $\lambda_p^{(q)}=V_{pb}V^*_{pq}$,
one observes that for certain processes the structure of the IR
divergences at NLO in QCDF is the same for both pieces. This allows to identify an
IR safe quantity {\bf at this order}, defined by $\Delta=T-P$ that can be evaluated safely in QCDF and that will be taken as the main input from QCDF. Another important remark is that this quantity can be directly related to observables leading to a set of sum rules that can be translated into predictions for
the UT angles\cite{dmv2,acta,alp} (see Strategy 2 in Sec.3.1).
In \cite{dmv2} this idea was extended to vector-vector final states. In this case,
there is a $\Delta$ associated to each helicity amplitude. But we focus on the leading (in a naive power counting in $\Lambda/m_b$)
longitudinal one. We obtain for the longitudinal $\Delta$ of the
$B_d\to K^{*0}\bar{K}^{*0}$ ($B_s\to K^{*0}\bar{K}^{*0}$)
decay
denoted  by $\Delta^{d}_{K^*K^*}$ ($\Delta^{s}_{K^*K^*}$)\cite{dmv2}:
\begin{eqnarray}
|\Delta^d_{K^*K^*}|&\!=\!&A_{K^*K^*}^{d,0} \frac{C_F \alpha_s}{4\pi N_c}C_1\,|\bar{G}_{K^*}(s_c)-\bar{G}_{K^*}(0)|\nonumber\\&\!=\!&(1.85 \pm 0.79)\times 10^{-7}\ {\rm GeV}\nonumber\\
|\Delta^s_{K^*K^*}|&\!=\!&A_{K^*K^*}^{s,0} \frac{C_F \alpha_s}{4\pi N_c}C_1\,|\bar{G}_{K^*}(s_c)-\bar{G}_{K^*}(0)|\nonumber\\&\!=\!&(1.62 \pm 0.69)\times 10^{-7}\ {\rm GeV} \label{equdeltas}
\end{eqnarray}
where $\bar{G}_V\equiv G_V-r_{\chi}^V \hat{G}_V$ are the usual
penguin functions and $A_{V_1 V_2}^{q,0}$ are the naive
factorization factors. The corresponding $\Delta$'s for the other modes can be found in \cite{dmv2}.

In short the method consist in relating the hadronic complex parameters $P_s^C$, $T_s^C$  that describes the dynamics of a $b \to s$ governed transition with the corresponding parameters of an U-spin $b \to d$ related process. This requires to include SU(3) breaking factorizable ($f=A_{V_1 V_2}^s/A_{V_1 V_2}^d$) and non-factorizable U-spin breaking $1/m_b$ suppressed corrections. Those non-factorizable corrections are sensitive to the different distribution amplitude of a $B_d$ and $B_s$ and spectator quark dependent contributions coming from gluons emitted from a $d$ or $s$ quark.
The next step is to determine the complex hadronic parameters of the $b \to d$ related decay $P_d^C$, $T_d^C$. This is done using the data on BR and direct CP asymmetry of the $b \to d$ decay and its associated $\Delta=T_d^{C}-P_d^{C}$ computed in QCDF.

The method was first applied to $B_s \to KK$ decays in the SM\cite{dmv} and supersymmetry \cite{bskksusy},
leading to the most precise predictions for $B_s \to K^+K^-$ and $B_s \to K^0 {\bar K^0}$ decay modes.

\subsection{$B_s \to VV$: A way to extract $\phi_s$.}
\medskip
Recent controversial claims on evidence of New Physics in the weak mixing phase $\phi_s$ \cite{ciusil} has focused the attention into this mixing phase. Here I will describe three possible strategies to measure this phase using $B$ mesons decaying into vectors that were discussed in \cite{dmv2}. We will focus on the golden mode $B_s \to K^{0*} {\bar K^{0*}}$ which can be easily reconstructed from the decays of the $K^{*}$ into kaons and pions.

{\it \underline{First Strategy}:} It applies to $B_s \to K^{0*} {\bar K^{0*}}$, $B_s \to \phi {\bar K ^{*0}}$ and $B_s \to \phi\phi$ decays.
 This strategy requires to measure the longitudinal BR and mixing-induced CP asymmetry of those modes and compute its corresponding $\Delta=T-P$ (see Eq.\ref{equdeltas}) from QCDF.

Expanding the longitudinal mixing-induced CP asymmetry in power of $\lambda_u^{s}/\lambda_c^{s}$ one obtains:
\begin{equation}
\label{pr}\Amix\lg(B_s\to K^{*0} \bar{K}^{*0})\simeq \sin{\phi_s}+\Delta S
\end{equation}
with $\Delta S=2\left|
\frac{\lambda_u^{(s)}}{\lambda_c^{(s)}} \right|
{\rm Re}\left( \frac{T^s_{K^*K^*}}{P^s_{K^*K^*}} \right)
 \sin{\gamma}\cos{\phi_s}+\cdots$
In order to evaluate the size of the $\Delta S$ pollution, one must constrain the size of ${\rm Re}(T/P)$
 and  translate these constrains into bounds on $\Delta S$. These bounds can be found in \cite{dmv2}.
The steps to follow are: first, one measures the longitudinal $BR(B_s \to K^{0*} {\bar K^{0*}})$,  second, for each value of this BR and using the bounds on ${\rm Re}(T/P)$ one obtains a possible range for $\Delta S $ \cite{dmv2}. Finally, once measured
 $A_{mix}^{long}$ of this decay mode one can then determine a range for $\sin \phi_s$ from:
$$\nonumber
\big(\Amix\lg-\Delta S_{max}\big)\ < \sin{\phi_s} <\ \big(\Amix\lg-\Delta S_{min}\big)\nonumber
$$
If this range is inconsistent with the predicted SM value for $\phi_s$ that would signal the presence of New Physics.

{\it {\underline{Second Strategy:}}} It is quite general, it applies to any $B$ decay into two pseudoescalars or vectors. For what concerns the measurement of $\phi_s$ we are interested in, we will focus here on two cases:
\begin{enumerate}
\item $B_s$ decay through a $b\to s$ process, e.g. $B_{s}\to K^{*0} \bar{K}^{*0}$
\item $B_s$ decay through a $b\to d$ process, e.g. $B_{s}\to \phi
  \bar{K}^{*0}$
(with a subsequent decay into a CP eigenstate)
\end{enumerate}
The great advantage of this strategy is that by measuring the longitudinal branching ratio, and the direct and mixing induced CP asymmetry of a $B_s$ meson decaying through a $b \to d$ or $b \to s$ process one gets a direct determination of the weak mixing angle $\phi_s$ with only {\bf one} single theoretical input: the corresponding $\Delta$ of the process. Even more, the precise way the asymmetries enter into this expression tells you that a measurement of the branching ratio and the 'untagged rate' is enough to determine $\beta_s$ but also $\gamma$. This can be seen explicitly in the expressions\cite{dmv2,acta}:
\begin{equation}
\sin^2{\beta_s}=\frac{\widetilde{BR}}{2|\lambda_c^{(D)}|^2|\Delta|^2}\left(
1-A_{\Delta \Gamma} \right)
\end{equation}
\begin{equation}
\sin^2{\left(\beta_s+\gamma\right)}=
\frac{\widetilde{BR}}{2|\lambda_u^{(D)}|^2|\Delta|^2}\left( 1-A_{\Delta \Gamma} \right)
\end{equation}
where $A_{\Delta \Gamma}$ verifies $|A_{dir}|^2+|A_{mix}|^2+|A_{\Delta \Gamma}|^2=1$.
This extraction of $\beta_s$ is done in this strategy under the assumption that there is no significant New Physics affecting the $b \to s$ decay.
This strategy has the advantage of minimizing the theoretical input, but it requires to  measure
several of the $B_s$ observables.

{\it \underline{Third Strategy}}: This last strategy is the most theoretically driven. It focus only on the golden mode $B_s \to K^{0*} {\bar K^{0*}}$ and assumes no sizeable New Physics in the $B_d \to K^{0*} {\bar K^{0*}}$ U-spin related decay.
The steps to follow here are the same that were done for $B\to KK$ decays.
First, one has to relate the hadronic parameters of both processes:
\begin{equation}
P_{K^*K^*}^{s}= f\,P_{K^*K^*}^{d}(1+\delta_{\sss K^*K^*}^P)\end{equation}
\begin{equation}
T_{K^*K^*}^{s}= f\,T_{K^*K^*}^{d}(1+\delta_{\sss K^*K^*}^T)
\label{TsPs}
\end{equation}
computing factorizable
$$f={m_{B_s}^2A_0^{B_s\to K^*}}/{m_B^2A_0^{B\to K^*}} = 0.88\pm 0.19$$
and non-factorizable SU(3) breaking parameters:
$$
|\delta_{\sss K^*K^*}^P|\le 0.12\ , \qquad |\delta_{\sss K^*K^*}^T|\le 0.15
$$
%
\begin{figure}[htb]\label{amixphis}
\includegraphics[height=12pc,width=14pc]{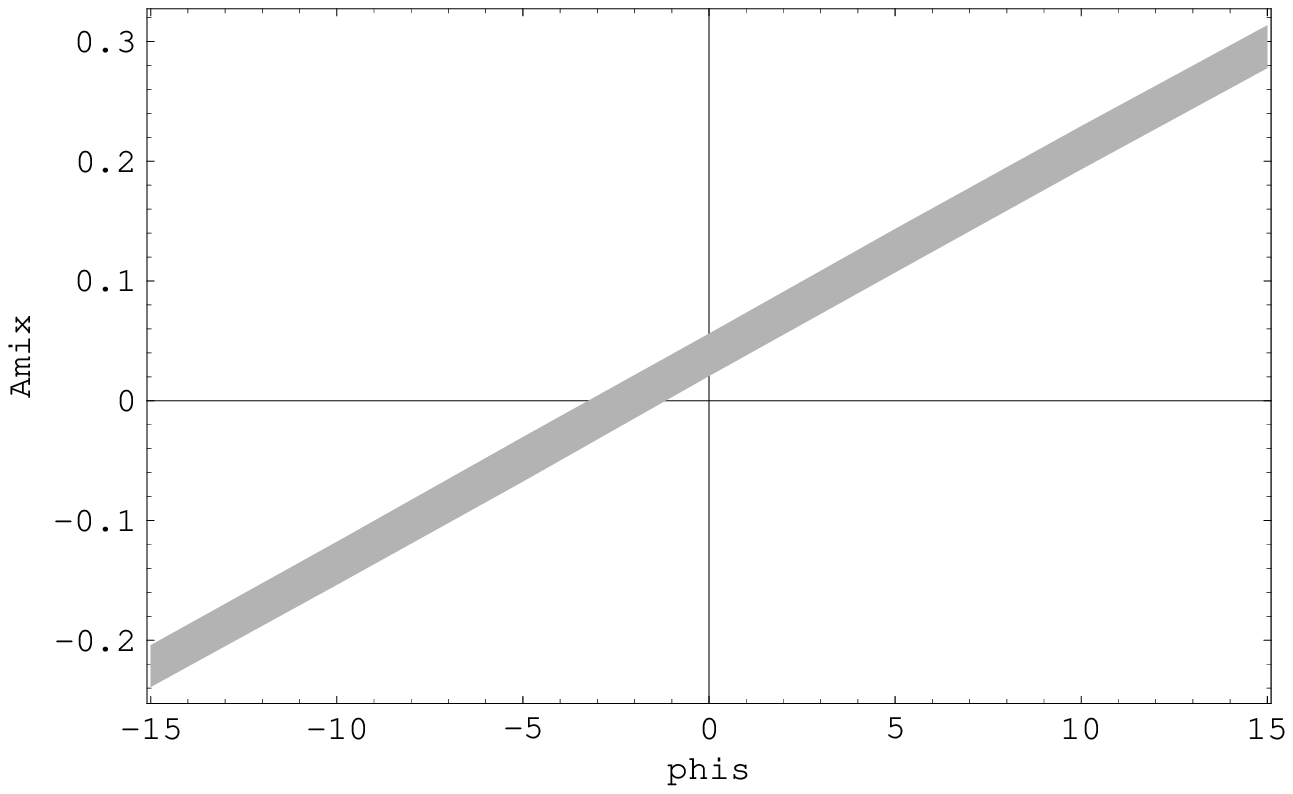}
\vspace*{-1cm}
\caption{${\cal A}_{mix}^{long}(B_s \to K^{0*} {\bar K^{0*}})$ versus $\phi_s$.}
\label{fig:largenenough}
\end{figure}

Then using as main inputs the  $BR^{long}(B_d \to K^{0*} {\bar K^{0*}})$, $\Delta_{K* K*}$ together with the longitudinal direct CP asymmetry of $B_d \to K^{0*} {\bar K^{0*}}$ one obtains a prediction in the SM for the corresponding $B_s$ observables: \begin{eqnarray}
\displaystyle
\left(\frac{BR\lg(B_s\to K^{*0}\bar{K}^{*0})}{BR\lg(B_d\to K^{*0}\bar{
K}^{*0})}\right)_{\sss }                &=&  17\pm 6 \nonumber\\
 \Adir\lg(B_s\to K^{*0}\bar{K}^{*0})_{\sss }
  &=& 0.000\pm 0.014 \nonumber  \\
\Amix\lg(B_s\to K^{*0}\bar{K}^{*0})_{\sss }               &=& 0.004 \pm 0.018
\nonumber
\end{eqnarray}
Finally a measurement of the longitudinal mixing induced CP asymmetry of $B_s \to K^{0*} {\bar K^{0*}}$ allow to extract the weak mixing angle $\phi_s$ including all penguin pollution. Figure 5 shows the correlation between $A_{mix}^{long}$ and $\phi_s$. The extraction of $\phi_s$ from this plot
is possible even in the presence of New Physics under the condition that there are only New Physics contributions in $\Delta B=2$ but not  large New Physics effects in $\Delta B=1$ FCNC amplitudes.
This requirement can be easily accomplished for generic type of New Physics models if two conditions are fulfilled (see \cite{fim}): i) $\Lambda_{eff}^{NP} \ll \Lambda_{ew}$ and ii) the effective coupling in $\Delta B=2$ can be expressed as the square of the effective coupling in $\Delta B=1$ amplitudes.
These conditions can be easily understood using an effective lagrangian language\cite{fim}.



\noindent {\it Summary:} Three different strategies to extract the weak mixing phase $\phi_s$ from $B \to VV$ were presented. Also the recent results on the $B\to K^*l^+l^-$ decay are discussed.\\

\noindent {\it Acknowledgements}: I would like to thank Giulia for such a nice and pleasant conference. I acknowledge financial support from FPA2005-02211, 2005-SGR-00994 and RyC program.

\end{document}